\documentclass{emulateapj}
\newcommand{\LCDM}{$\Lambda$CDM}
\newcommand{\msun}{\mbox{${\rm M}_{\odot}$}}
\newcommand{\sersic}{S\'ersic}

\newcommand{\gems}{{GEMS}}

\newcommand{\vcir}{\mbox{$V_c$}}
\newcommand{\Jd}{\mbox{$J_d$}}
\newcommand{\rd}{\mbox{$r_d$}}

\newcommand{\mvir}{\mbox{$M_{\rm vir}$}}
\newcommand{\rvir}{\mbox{$r_{\rm vir}$}}
\newcommand{\vvir}{\mbox{$V_{\rm vir}$}}
\newcommand{\vdisk}{\mbox{$V_{\rm disk}$}}
\newcommand{\cvir}{\mbox{$c_{\rm vir}$}}

\shorttitle{An Explanation for the Observed Weak Size Evolution of Disk Galaxies}

\begin{document}

\title{An Explanation for the Observed Weak Size Evolution of Disk Galaxies}

\shortauthors{Somerville et al.}

\author{
Rachel S.~Somerville\altaffilmark{1}, 
Marco Barden\altaffilmark{1}, 
Hans-Walter Rix\altaffilmark{1}, 
Eric F.~Bell\altaffilmark{1}, 
Steven V.~W.~Beckwith\altaffilmark{3,4}, 
Andrea Borch\altaffilmark{1,2}, 
John A.~R.~Caldwell\altaffilmark{5}, 
Boris H\"au\ss ler\altaffilmark{1}, 
Catherine Heymans\altaffilmark{6}, 
Knud Jahnke\altaffilmark{1}, 
Shardha Jogee\altaffilmark{7}, 
Daniel H.~McIntosh\altaffilmark{8}, 
Klaus Meisenheimer\altaffilmark{1}, 
Chien Y.~Peng\altaffilmark{3}, 
Sebastian F.~S\'anchez\altaffilmark{9}, 
Lutz Wisotzki\altaffilmark{10}, 
Christian Wolf\altaffilmark{11}
}
\email{somerville@mpia.de}

\altaffiltext{1}{Max-Planck-Institut f\"ur Astronomie, K\"onigstuhl
  17, Heidelberg, 69117, Germany}
\altaffiltext{2}{Astronomisches Rechen-Institut, M\"onchhofstr. 12-14, D-69120, Heidelberg, Germany}
\altaffiltext{3}{Space Telescope Science Institute, 3700 San Martin Dr., Baltimore, MD 21218, USA}
\altaffiltext{4}{Johns Hopkins University, 3400 North Charles
  Street, Baltimore, MD 21218, USA}
\altaffiltext{5}{University of Texas, McDonald Observatory, Fort
  Davis, TX 79734}
\altaffiltext{6}{Department of Physics and Astronomy, University of British Columbia, 6224 Agricultural Road, Vancouver, V6T 1Z1, Canada}
\altaffiltext{7}{Department of Astronomy, University of Texas at Austin, 1 University Station, C1400 Austin, TX 78712-0259, USA}
\altaffiltext{8}{Department of Astronomy, University of Massachusetts, 
710 North Pleasant Street, Amherst, MA 01003, USA}
\altaffiltext{9}{Centro Hispano Aleman de Calar Alto, C/Jesus Durban
  Remon 2-2, Almeria, E-04004, Spain}
\altaffiltext{10}{Astrophysikalisches Institut Potsdam, An der
  Sternwarte 16, Potsdam, 14482, Germany}
\altaffiltext{11}{Department of Physics, Denys Wilkinson Bldg.,
  University of Oxford, Keble Road, Oxford, OX1 3RH, UK}

\begin{abstract}
Surveys of distant galaxies with the Hubble Space Telescope and from
the ground have shown that there is only mild evolution in the
relationship between radial size and stellar mass for galactic disks
from $z\sim1$ to the present day. Using a sample of nearby
disk-dominated galaxies from the Sloan Digital Sky Survey (SDSS), and
high redshift data from the GEMS (Galaxy Evolution from Morphology and
SEDs) survey, we investigate whether this result is consistent with
theoretical expectations within the hierarchical paradigm of structure
formation. The relationship between virial radius and mass for dark
matter halos in the \LCDM\ model evolves by about a factor of two over
this interval. However, N-body simulations have shown that halos of a
given mass have less centrally concentrated mass profiles at high
redshift. When we compute the expected disk size-stellar mass
distribution, accounting for this evolution in the internal structure
of dark matter halos and the adiabatic contraction of the dark matter
by the self-gravity of the collapsing baryons, we find that the
predicted evolution in the mean size at fixed stellar mass since
$z\sim1$ is about 15--20 percent, in good agreement with the
observational constraints from GEMS. At redshift $z\sim2$, the model
predicts that disks at fixed stellar mass were on average only 60\% as
large as they are today. Similarly, we predict that the rotation
velocity at a given stellar mass (essentially the zero-point of the
Tully-Fisher relation) is only about 10 percent larger at $z\sim1$ (20
percent at $z\sim2$) than at the present day.
\end{abstract}

\keywords{
galaxies: spiral -- galaxies: evolution -- galaxies: 
high redshift -- surveys -- cosmology: observations
}

\section{Introduction}
\label{sec:intro}

The relationship between the radial size and the luminosity or stellar
mass of galactic disks is a fundamental scaling relation that reveals
important aspects of the formation history of these objects. The size,
luminosity/mass, and rotation velocity form a `fundamental plane' for
disks at the present epoch \citep{pizagno:06} that is analogous to the
more familiar fundamental plane for early-type galaxies
\citep{burstein:97}. The zero-point, slope, and scatter of the
fundamental plane for both disks and spheroids, and the evolution of
these quantities over cosmic time, pose strong constraints on models
of galaxy formation.

In the modern paradigm of disk formation, Cold Dark Matter (CDM)
dominates the initial gravitational potential, and dark matter (DM)
and gas aquire angular momentum via tidal torques in the early
universe \citep{peebles:69}. When the gas cools and condenses, this
angular momentum may eventually halt the collapse and lead to the
formation of a rotationally supported disk
\citep{fall_efstathiou:80}. Under the assumption that the specific
angular momentum of the pre-collapse gas is similar to that of the DM
and is mostly conserved during collapse, this picture leads to
predictions of present-day disk sizes that are in reasonably good
agreement with observations
\citep{kauffmann:96,dalcanton:97,mo:98,afh:98,somerville_primack:99,vandenbosch:00,dutton:07,gnedin:06}.

However, in the most sophisticated numerical hydrodynamic simulations
of disk formation in a CDM universe, the proto-disk gas tends to lose
a large fraction of its angular momentum via mergers, leading to disks
that are too small and compact
\citep{navarro_white:94,sommer_larson:99,navarro_steinmetz:00}. It is
still not clear whether this problem reflects a fundamental difficulty
with CDM (i.e., excess small scale power), or is due to inadequate
numerical resolution or treatment of ``gastrophysical'' processes like
star formation and feedback
\citep{governato:04,robertson:04}. However, it has been suggested that
delayed cooling and star formation, perhaps due to strong feedback in
low-mass progenitors, could help to stem this angular momentum loss
\citep{weil:98,maller-dekel:02}. These ideas can be tested by
observing the evolution of disk scaling relations at high redshift
relative to the present epoch.

The observational relationship between radial size (effective radius
or disk scale length) and luminosity or stellar mass for disks at low
redshift has now been well-characterized by studies based on the Sloan
Digital Sky Survey \citep[SDSS; e.g.][hereafter S03]{shen:03}. Several
pioneering studies in the past decade studied the size-luminosity
relation for disks out to redshift $z\sim1$
\citep[e.g.][]{lilly:98,simard:99}. Sizes were also measured for Lyman
Break Galaxies (LBGs) at redshifts $z\sim3$
\citep{giavalisco:96,lowenthal:97}. However, there was significant
disagreement between the results of different studies, and a clear
understanding of the evolution of galaxy sizes over cosmic time was
hampered by the difficulty of obtaining large samples with accurate
redshifts and high-resolution imaging, and by concerns about the
impact of surface brightness selection effects. An additional problem
with interpreting high redshift data is that as higher redshifts are
probed, the observed optical begins to shift into the rest-UV, and
k-corrections become highly uncertain. Moreover, the stellar
mass-to-light ratios of galaxies increase as their stellar populations
age, and therefore the size-\emph{luminosity} relation would change
over time even for galaxies that were just ``passively'' aging.

Recently, new studies with the Advanced Camera for Surveys (ACS) on
the Hubble Space Telescope (HST) have provided greatly improved
constraints on the disk size-luminosity relation at high
redshift. \citet{ravindranath:04} and \citet{ferguson:04} presented
size distributions for $z\la1$ disk-type galaxies and for rest-UV
selected galaxies from $1.4 \la z \la 6$, respectively, based on
samples selected from the Great Observatories Origins Deep Survey
(GOODS). \citet[][hereafter B05]{barden:05} presented the
luminosity-size and stellar mass-size relation out to $z\sim1$ based
on the GEMS (Galaxy Evolution from Morphology and SEDs) survey
\citep{rix:04}. They concluded that disk galaxies of a given size at
$z\sim1$ are $\sim$1 magnitude brighter in the $V$-band, but that
there is less than about a ten percent change in the \emph{stellar
  mass} at a given size between $z\sim1$ and the present.  This result
is consistent with a mean stellar mass-to-light ratio that increases
with time, as expected based on the simple aging of stellar
populations. An interesting complementary result was recently found by
\citet{sargent:06} based on the COSMOS survey. They found that the
number density of disks with half-light radii between 5 and 7 kpc is
nearly constant from $z\sim1$ until the present. Coupled with findings
that the stellar mass function of spiral galaxies at $z\sim1$ was
about the same as it is today \citep{borch:06}, this reinforces a
picture in which disks have a fixed relationship between their stellar
mass and their radial size over this redshift interval (i.e., that as
disks grow in mass through star formation, their sizes grow in such a
way as to keep them on this relation, on average).

\citet{trujillo:04} measured luminosity-size and stellar mass-size
relations in the rest-frame optical (based on the ground-based Near
Infrared selected FIRES sample) out to $z\sim2.5$, and found that the
average surface brightness at $z\sim2.5$ is about 2--3 mag
arcsec$^{-1}$ brighter than in the local universe, but the average
size at a fixed stellar mass has evolved by less than a factor of
two. \citet[][hereafter T06]{trujillo:06} presented the results of a
similar analysis of a larger sample from FIRES, and combined those
results with the lower redshift studies of S03 and B05.

How do these observations compare with theoretical expectations? In
the simplest version of the Fall-Efstathiou picture, under the
assumption that halo mass density profiles are singular isothermal
spheres (SIS; $\rho(r) \propto 1/r^2$) and neglecting the self-gravity
of the baryons, we expect the size of a galactic disk that forms
within a dark matter halo to scale as $r_{\rm disk} \propto \lambda
r_{\rm vir}$, where $\lambda$ is the dimensionless spin parameter and
$r_{\rm vir}$ is the virial radius of the dark matter halo. N-body
simulations have demonstrated that the spins of dark matter halos are
not correlated with halo mass or most other properties, and the
distribution does not evolve with time \citep[e.g.][]{bullock:01b}. If
the stellar mass of the disk is a constant fraction of the halo virial
mass, then, in this simple picture, the expectation is that the
average size of galactic disks of a given stellar mass will evolve as
$r_{\rm vir}$ evolves for halos of a given virial mass. In the
currently favored \LCDM\ cosmology, this would imply a decrease of a
factor of $\sim 1.7$ out to $z=1$ and a factor of $\sim3$ out to
$z\sim3$. This $r_{\rm disk} \propto \lambda r_{\rm vir}$ scaling
(hereafter referred to as the SIS model) has frequently been used in
the literature as a theoretical baseline
\citep[e.g.][B05,T06]{mao:98,ferguson:04}. \citet{mao:98} found that
the SIS model was consistent with the size evolution of disks out to
$z\sim1$ compared with the data available at the time, but the samples
were tiny, and the observational selection effects were not well
characterized or accounted for. \citet{ferguson:04} also found that
the average rest-UV sizes of rest-UV selected galaxies at $1.4\la z
\la 5$ were consistent with the SIS model, but the connection of
rest-UV luminosity with stellar mass (or halo mass) is quite
uncertain. Most recently, B05 and T06 concluded that the predicted
evolution in the SIS model is considerably stronger than the observed
evolution of the rest-optical sizes in their stellar-mass selected
disk samples.

However, the SIS model neglects several important factors that are
believed to play a role in determining the size of galactic disks
forming in CDM halos. 1) The mass density profiles of CDM halos are
not SIS, but have a universal form \citep[known as the
  Navarro-Frenk-White (NFW) profile;][]{navarro:97}, characterized by
the concentration parameter $\cvir$. The concentration parameter
quantifies the density of the halo on small ($\sim$ kpc) scales
relative to the virial radius, and has an important impact on the
structural parameters of the resulting disk. There is a correlation
between halo virial mass $M_{\rm vir}$ and concentration
\citep{navarro:97}, though with a significant scatter
\citep{bullock:01a}, and this mean halo concentration-mass relation
evolves with time, in the sense that halos of a given mass were less
concentrated in the past \citep{bullock:01a}. 2) The self-gravity of
the baryonic material may modify the distribution of the dark matter
as it becomes condensed in the central part of the halo (``adiabatic
contraction'') 3) Disks with low values of $\lambda$ and/or large
baryonic-to-dark mass ratios may not have sufficient angular momentum
to support a stable disk. These unstable disks may form a bar or a
bulge, and might no longer be included in a sample of `disk dominated'
galaxies.

There are of course numerous other potential complications in the
process of the formation and evolution of galactic disks, which we do
not consider here (though we discuss some of them in
\S\ref{sec:conclude}). Here we present the predictions of a model for
disk formation that improves on the simple SIS model by incorporating
NFW halo profiles, adiabatic contraction, and disk instability. This
model is based on the formalism and basic ingredients presented in
\citet{blumenthal:86}, \citet{flores:93}, and
\citet[][MMW98]{mo:98}. Our main result is that the predictions of
this improved model are compatible with the rather weak observed
evolution of the disk stellar mass-size relation out to $z\sim 1$
reported by B05. We also extend these predictions out to higher
redshift $z\sim3$, and find acceptable agreement with the results
reported by T06.

We discuss the ingredients of our model in \S\ref{sec:model}, give a
brief summary of the observational data in \S\ref{sec:data}, present
our results in \S\ref{sec:results}, and discuss our results and
conclude in \S\ref{sec:conclude}. We assume the following values for
the cosmological parameters: matter density $\Omega_m = 0.3$, baryon
density $\Omega_b =0.044$, cosmological constant
$\Omega_{\Lambda}=0.70$, Hubble parameter
$H_0=70$\,km\,s$^{-1}$\,Mpc$^{-1}$, fluctuation amplitude $\sigma_8 =
0.9$, and a scale-free primordial power spectrum $n_s=1$.

\section{Model}
\label{sec:model}

The fundamental hypothesis of our model, following the Fall-Efstathiou
paradigm, is that galactic disks form within massive, extended dark
matter halos.  The structural properties and multiplicity functions of
dark matter halos in a given cosmology can be robustly predicted from
N-body simulations and analytic fitting formulae. Therefore, the main
challenge is to relate the structural properties of dark matter halos
to the observable properties of disk galaxies. Making this connection
is the goal of our model, which we now briefly present. Note that our
approach, and the exposition here, are closely based on the formalism
presented in \citet{blumenthal:86}, \citet{flores:93}, and
\citet[][MMW98]{mo:98}.

We assume that a fraction $f_d \equiv m_d/\mvir$ of the halo's virial
mass is in the form of baryons that are able to cool and collapse,
forming a disk with mass $m_d$. We denote the angular momentum of the
disk material as $J_d$, and assume that it is a fixed fraction of the
halo angular momentum $J_h$ and that specific angular momentum is
conserved when the baryons cool and collapse. We also assume that the
disk is thin, in centrifugal balance, and has an exponential surface
density profile $\Sigma (r)=\Sigma_0 \exp\left(-r/r_d\right)$.  Here
$r_d$ and $\Sigma_0$ are the disk scalelength and central surface
density, and are related to the disk mass through $m_d = 2\pi \Sigma_0
{r_d}^2$.

The angular momentum of the disk is: 

\begin{equation} \label{jdisk}
\Jd=2\pi \int \vcir(r) \Sigma (r) r^2  d r.
\end{equation}

If we were to assume that the initial dark matter density profile is a
singular isothermal sphere, and to neglect the self-gravity of the
disk, then the rotation velocity $V_c(r)$ is constant and equal to
$V_{\rm vir}$, and the angular momentum of the disk reduces to:

\begin{equation} \label{eqn:jdisk}
J_d =4\pi \Sigma_0 \vcir \rd^3 = 2m_d r_d\vvir
\end{equation}

Using the definition of the spin parameter, $\lambda \equiv J_h\vert
E_h\vert ^{1/2} G^{-1} \mvir^{-5/2}$ \citep{peebles:69}, where $E_h$
is the total energy of the halo, we can write:

\begin{equation} \label{eqn:rdisk_sis1}
\rd={\lambda G\mvir ^{3/2}\over 2\vvir \vert E_h \vert ^{1/2}}
\left[{(\Jd/m_d) \over (J_h/\mvir)}\right].
\end{equation}

We now define $f_j \equiv (J_d/m_d)/(J_h/\mvir)$ to be the ratio of
the \emph{specific} angular momenta of the disk and the halo. 

If we assume all particles to be on circular orbits, the total energy
of a truncated singular isothermal sphere can be obtained from the
virial theorem:

\begin{equation}
E_h = -{G\mvir^2\over 2\rvir} = -{\mvir\vvir^2\over 2}.
\end{equation}

Inserting this expression in eqn.~\ref{eqn:rdisk_sis1}, we obtain: 
\begin{equation}\label{eqn:rdisk_sis2}
r_d ={1\over \sqrt {2}} f_j \lambda \rvir
\end{equation}.

If we assume that the spin parameter does not change with cosmic
epoch, this leads to the expected scaling $r_d(z) \propto
\rvir(z)$. We refer to this model as the `SIS' model, reflecting the
assumption that the initial halo profiles are singular isothermal
spheres.

A more realistic model would incorporate cosmologically motivated
Navarro-Frenk-White (NFW) halo profiles \citep{navarro:97} and should
also include the effect of the self-gravity of the baryons on the
predicted properties of the disk. If the gravitational effects of the
disk were negligible, then the rotation curve would be the same as
that of the unperturbed NFW halo, which rises to a maximum value
$V_{\rm max}$ at a radius $\sim 2r_s$, where $r_s$ is the NFW scale
radius $r_s \equiv \rvir/\cvir$. However, on scales where the
mass of baryonic material becomes significant compared with the dark
matter, we must account not only for the direct gravitational effects
of the baryonic disk material, but also for the fact that the dark
matter contracts in response to the gravitational force of the baryons
as they collapse to form the disk. We compute the effect of this
contraction under the assumption that the disk forms gradually enough
that the response of the halo is ``adiabatic'', i.e., that the
following ``adiabatic invariant'' quantity is conserved:
\begin{equation}
GM_f(<r_f)r_f = GM_i(<r_i) r_i \, ,
\end{equation}
where $M_i(<r_i)$ is the initial mass within an initial radius $r_i$
(assumed to be given by the NFW profile), and $M_f(<r_f)$ is the
post-collapse mass within a final radius $r_f$. Numerical tests of the
validity of the ``adiabatic collapse'' formalism have shown that it
works surprisingly well, even when some of the underlying assumptions
(such as spherical symmetry and all particles being on circular
orbits) are violated \citep{jesseit:02,gnedin:04}.

The final mass within a radius $r$ is the mass of the exponential disk
plus the mass of dark matter within the initial radius $r_i$: $M_f(r)
= m_d(r) + M(r_i)(1-f_d)$. The modified rotation curve can then be
written as a sum in quadrature of the contributions from the disk and
the (contracted) dark matter halo: ${V_c}^2(r) = {V_{c,d}}^2(r) +
{V_{c,DM}}^2(r)$. The rotation curve for the disk is given by the
usual expression for a thin exponential disk \citep[see
  e.g.][]{binney_tremaine:87}, and the dark matter component of the
rotation velocity is given by ${V_{c,DM}}^2(r)=G[M_f(r)-m_d(r)]/r$.
Using eqn.~\ref{eqn:jdisk} again, and making a change in variable
$u=r/r_d$, we can write the expression for the angular momentum of the
disk as:
\begin{equation}
J_d = 
m_d \rd \vvir \int_0^{\rvir/r_d}
e^{-u} u^2 {\vcir(\rd u) \over \vvir~} du,
\end{equation}
and, in direct analogy to eqn.~\ref{eqn:rdisk_sis2}, we obtain
\begin{equation} \label{rd_real}
r_d = {1\over \sqrt{2}} f_j 
\lambda \rvir f_c^{-1/2} f_R (\lambda,c,f_d)
\end{equation}
where 
\begin{equation}
f_R (\lambda,c,f_d) 
= 2 \left[\int_0^\infty
e^{-u} u^2 ~{\vcir(\rd u)\over\vvir}~ du\right]^{-1}.
\end{equation}
and $f_c \equiv E_{\rm NFW}/E_{\rm SIS}$ is the total energy for a
halo with an NFW profile, relative to that for a singular isothermal
sphere with the same mass. Note that in this step we have also used
the fact that $\rvir >> r_d$, and that the disk density profile
declines exponentially, to set the upper limit of integration to
infinity.

For a given set of halo properties \mvir, \cvir, and $\lambda$, and
disk parameters $f_d$ and $f_j$, we can solve this system of equations
iteratively to obtain $\vcir(r)$ and $r_d$ \footnote{For a more
  detailed account of how to go about solving the equations, and
  analytic fitting functions for $f_R$ and $f_c$, see MMW98)}. We
describe how we obtain the halo properties and disk parameters in the
following section. We will refer to the improved model as the `NFW'
model.

\begin{figure*} 
\begin{center}
\epsscale{1.0}
\plotone{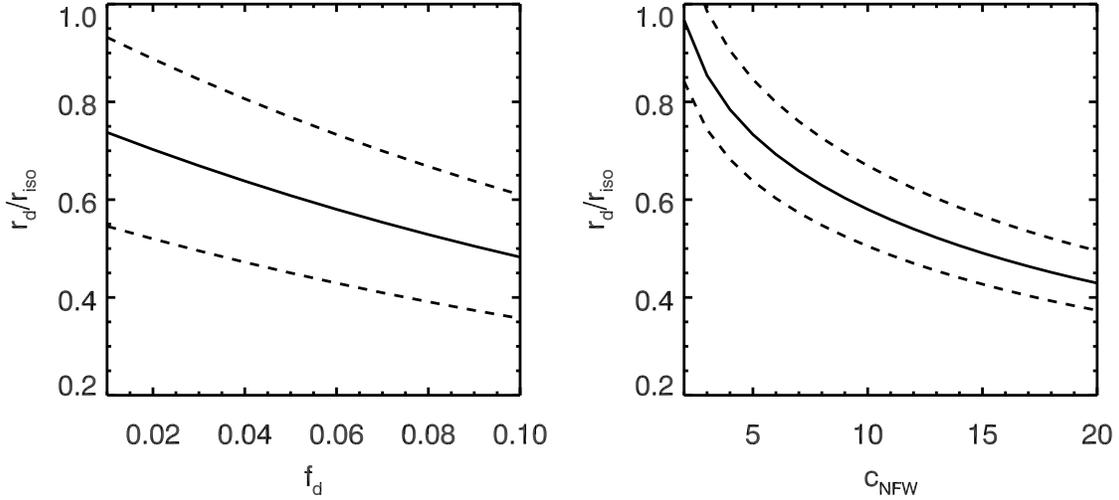}
\end{center}
\caption{\small The ratio of the disk scale length predicted in the
  NFW model with adiabatic contraction, to that in the SIS model
  (neglecting disk self-gravity), as a function of the fraction of
  baryons in the disk $f_d$ (left panel) and the NFW halo
  concentration $\cvir$. For all curves, the mean value of the spin
  parameter $\lambda=0.05$ has been assumed. In the left panel, the
  solid curves are for \cvir=10, while the upper and lower dashed
  curves are for \cvir=5 and 20, respectively. In the right panel, the
  solid curve is for $f_d=0.06$ and the upper and lower curves are for
  $f_d=0.03$ and $f_d=0.09$, respectively. 
\label{fig:rd_riso}}
\epsscale{1.0}
\end{figure*}

Comparing eqn.~\ref{rd_real} with the corresponding expression in the
SIS model, eqn.~\ref{eqn:rdisk_sis2}, we can see that they differ by
two factors: $f_c^{-1/2}$, which reflects the difference in energy of
a singular isothermal sphere vs. a NFW profile, and $f_R$, which
reflects both the NFW density profile and the effect of the adiabatic
contraction. The ratio of eqn.~\ref{rd_real} and
eqn.~\ref{eqn:rdisk_sis2} ($r_d$/$r_{iso}$) is plotted in
Fig.~\ref{fig:rd_riso}, as a function of disk fraction $f_d$ and of
halo concentration $\cvir$. From this plot, we can gain several
insights into the behavior of the model. First, we can see that the
NFW model generally predicts disk sizes that are smaller than the
values in the SIS model. For a dark matter halo with
$\mvir=10^{12}\msun$, the present-day virial radius is $\rvir \simeq
260$ kpc, and so from eqn~\ref{eqn:rdisk_sis2}, the predicted scale
radius would be about 6 kpc for the median value of the spin parameter
$\lambda=0.035$. This is almost a factor of two larger than one might
expect, if such a halo hosts a galaxy similar to the Milky Way or
M31. Therefore, there is room for this reduction in size predicted by
the new model.

We can also see that smaller values of $f_d$ lead to \emph{larger}
values of $r_d$. This is because there is less mass and therefore less
gravity to contract the halo, leading to a more extended
disk. Finally, we see that lower values of concentration \cvir\ also
lead to larger (more extended) disks. Lower values of \cvir\ imply
that there is less mass near the center of the halo relative to the
outer parts, again leading to weaker gravitational forces on the kpc
scales where the disk is forming, and less contraction of the halo,
therefore a more extended disk.

\subsection{Disk Stability}

Because of the rather large range of allowed spin parameters
$\lambda$, our model predicts a fairly broad distribution of galaxy
sizes at fixed stellar mass. However, it is possible that not all of
these disks gravitationally stable. It is well-known from numerical
studies that galaxies in which the self-gravity of the baryons
dominates may be unstable to the formation of a bar
\citep[e.g.][]{christodoulou:95}. Following MMW98, who base their
stability criterion on the study of \citet{efstathiou:82}, we adopt a
threshold for instability $\epsilon_m \equiv V_{\rm
  max}/(Gm_d/r_d)^{1/2} < \epsilon_{\rm m, crit}$, where $V_{\rm max}$
is the maximum rotation velocity, which we approximate as the rotation
velocity at $\sim 3 r_d$. We adopt $\epsilon_{\rm m, crit}=0.75$
\citep{syer:99}, and assume that galaxies with $\epsilon_m$ less than
this critical value may form a significant bar or bulge and be
excluded from a disk sample. Because of the uncertainties associated
with the fate of these ``unstable'' disks, however, we present the
results of our analysis both including and excluding these objects.

\subsection{Initial Conditions and Model Parameters}

Let us consider a virialized dark matter halo of a given virial mass
($M_{\rm vir}$) at a specific redshift. We require the following
information, all as a function of redshift:

\begin{itemize} 

\item The number density of these halos

\item The halo virial radius and virial velocity

\item The initial halo mass density profile $\rho(r)$, here
  characterized by the Navarro-Frenk-White concentration parameter
  $\cvir$

\item The halo specific angular momentum, characterized by the
  dimensionless spin parameter $\lambda$.

\end{itemize}

All of this information can be robustly obtained with a high degree of
accuracy from modern numerical cosmological N-body simulations. We now
describe each ingredient in some detail, and how we obtain it.

\noindent \emph{Halo number density as a function of redshift:} Over a
given redshift interval, we select halo masses from the mass function
of \citet{sheth_tormen:99} using a Monte Carlo procedure.

\noindent \emph{Halo virial radius and virial velocity}: It is
standard to define dark matter halos as spherical regions within which
the average overdensity exceeds a critical value $\Delta_{\rm vir}$
times the mean background density of the universe. The value of
$\Delta_{\rm vir}$ is computed based on the spherical collapse
model. In an Einstein-de Sitter ($\Omega = 1$) universe, $\Delta_{\rm
  vir}=178$ and by definition the background density $\rho_{\rm b}$
equals the critical density $\rho_{\rm crit}$. During the early years
of the development of the CDM paradigm, when the Einstein-de Sitter
model was widely assumed to be correct, it became common to define
halos as overdensities greater than 200 times the \emph{critical}
density (the value 200 was obtained by just rounding off 178). This
assumption is still commonly used by many authors today, in spite of
the fact that it is no longer well-motivated in the now-favored
concordance cosmology. We instead use the appropriate value of
$\Delta_{\rm vir}$ for our adopted cosmology, as predicted by the
spherical collapse model ($\Delta_{\rm vir}(z=0) \simeq 337$) times
the \emph{mean} background density. This assumption is also commonly
used in the literature, and was used in the Bullock et al. (2001a,
2001b) papers from which we obtain our halo profile and spin
relations. The details of the adaptation of the spherical collapse
model to a general cosmology (with cosmological constant) and formulae
for and plots of the relations for halo virial radius as a function of
mass and redshift are given in the Appendix of
\citet{somerville_primack:99}, and are the same relations that we use
here. The halo virial velocity is then easily obtained by use of the
virial relation, $V_{\rm vir} = (GM_{\rm vir}/r_{\rm
  vir})^{1/2}$. Note that, because halos are defined with respect to
the background density of the universe, average halo densities were
higher in the past. This implies that halos of a given virial mass had
smaller virial radii, and higher virial velocities, in the past than
they do today.

\noindent \emph{Halo Density Profile}: We assume that initially the
halo mass density profiles obey the Navarro-Frenk-White form
\citep{navarro:97}, and compute the halo concentration $c_{\rm vir}$
as a function of mass and redshift using the analytic model provided
by \citet[][B01]{bullock:01a}, which is based on the results of
numerical simulations. The results of B01 have been confirmed by a
number of other studies \citep{jing:00,ens:01,wechsler:02,maccio:06}.

The basic ingredients of the B01 analytic model for $\cvir(M, z)$ are
as follows. They define a ``collapse redshift'' $z_c$ for each halo as
$M_*(z_c) = F M_{\rm vir}$, where $\sigma[M_*(z)] = \delta_c/D_{\rm
  lin}$. Here, F is a parameter of the model, which is measured by
fitting to N-body simulations, $\sigma(M)$ is the $z=0$ linear rms
density fluctuation on the comoving scale encompassing a mass $M$,
$\delta_c \simeq 1.686$ is the linear theory value of the critical
density in the spherical tophat model, and $D_{\rm lin}(z)$ is the
linear growth factor. Then the concentration is given by $c_{\rm
  vir}(M_{\rm vir}, z) = K (1+z_c)/(1+z)$, where $K$ is another
adjustable parameter of the model. Note that $z_c$ is independent of
$z$, and therefore $c_{\rm vir} \propto (1+z)^{-1}$ at fixed mass.

\citet{bullock:01a} found values of $F=0.01$ and $K=4$ for a
\LCDM\ model with the same values of $\Omega_m$, $\Omega_{\Lambda}$,
and $H_0$ as our cosmology, but with a higher value of $\sigma_8=1$. A
recent study by \citet{maccio:06}, based on a larger sample of higher
resolution simulations, confirms the B01 model but finds $K=3.4$ for a
\LCDM\ cosmology with $\sigma_8=0.9$ (corresponding to a 15 percent
lower normalization than B01, but with the same mass dependence).  In
the model presented here, we adopt the updated \citet{maccio:06}
normalization of the $c_{\rm vir}(\mvir)$ relation ($K=3.4$).

The concentration of halos of a given mass at a given redshift depends
on the cosmology and the power spectrum. In cosmologies in which halos
assemble their mass early (late), a halo of a given mass has a higher
(lower) concentration \citep{wechsler:02}. We can see from the above
formulae that in a cosmology with a modified $D_{\rm lin}(z)$ (for
example, if we changed the value of the cosmological constant,
$\Omega_\Lambda$), the redshift evolution of $c_{\rm vir}(\mvir)$
would be different. Similarly, if we modified the normalization or
shape of the power spectrum (which enters through $\sigma(M)$), it
would alter the normalization and slope of the $c_{\rm vir}(\mvir)$
relation. For example, in the cosmology derived from the WMAP three
year results of \citet{spergel:06}\footnote{$\Omega_m = 0.24$,
  $\Omega_b =0.042$, $\Omega_{\Lambda}=0.76$,
  $H_0=73$\,km\,s$^{-1}$\,Mpc$^{-1}$, $\sigma_8 = 0.74$, $n_s=0.95$},
the $c_{\rm vir}(\mvir)$ relation has a normalization about 30 \%
lower than in our adopted cosmology, but a nearly identical redshift
dependence.

\noindent \emph{Halo Specific Angular Momentum}: N-body simulations
have demonstrated that the spin parameter $\lambda$, which
characterizes the specific angular momentum of dark matter halos, is
uncorrelated with the halo's mass and concentration
\citep{bullock:01b,maccio:06} and does not evolve with redshift. The
distribution of $\lambda$ is log-normal, with mean
$\bar{\lambda}=0.05$ and width $\sigma_{\lambda}=0.5$
\citep{bullock:01b}. We therefore assign each halo a value of
$\lambda$ by selecting values randomly from this distribution,
assuming that it is not correlated with any other halo properties or
with redshift.
 
\subsection{Predicting observable disk properties}

We assume that each halo with a virial velocity below 350 km/s hosts
one disk galaxy at its center. Once the halo properties are specified
as described above (based on dissipationless N-body simulations),
there are only two free parameters in our model: the fraction of mass
in the form of baryons in the disk, $f_d$, and the fraction of the
specific angular momentum of the DM halo captured by the disk,
$f_j$. Throughout this work, we assume $f_j=1$. Although there are
physical reasons to think that $f_j$ may not be exactly equal to
unity, note that in order to change our main conclusions, $f_j$ would
have to be a strong function of redshift. There is no obvious physical
reason to expect this to be the case.

Since $f_d$ is now the only free parameter, we can fix its value by
requiring that we reproduce the observed scale length of a galaxy of a
given virial mass scale. For example, adopting $f_d=0.06$ produces a
galaxy with disk mass $m_d = 6 \times 10^{10} \msun$ and scale length
$r_d = 3.6$ kpc in a halo of mass $\mvir = 10^{12}\msun$, similar to
the Milky Way and in good agreement with the median value for the
scale length of disks in our SDSS sample at this stellar
mass. However, if we assume a constant value of $f_d=0.06$ for all
halo masses, we find that the slope of the stellar mass vs. disk scale
length relation is too steep, i.e. the sizes of smaller mass disks are
too small, and those of larger mass disks too large, compared with the
observed relation. This has also been noted before by other authors
working with similar models \citep[e.g.][]{shen:03,dutton:07}. We can
solve this problem by assuming that the disk baryon fraction varies
with halo mass, in the sense that lower mass halos form disks with
lower baryon fractions. This is in fact what is found in physically
motivated models of galaxy formation, as supernova feedback can more
easily heat and eject gas from low-mass halos
\citep{natarajan:99,dekel-silk:86}. We therefore adopt a simple
functional form for $f_d$ as a function of halo mass: $f_d = f_0
/[1.0+(\mvir/M_c)^{\alpha}]$, following S03, and find good agreement
with the observed stellar mass vs. disk scale length relation with
parameter values $f_0=0.13$, $M_c=1.0\times 10^{12} \msun$, and
$\alpha=-0.67$. Note that we do not discriminate between stellar mass
and the mass in cold gas. For the relatively massive disks that we
will focus on, the cold gas fraction should be fairly low.

\section{Summary of Observational Data}
\label{sec:data}

For the main part of our analysis, we use the same sample of
disk-dominated galaxies that was used for the analysis of B05. We give
a brief summary of that sample here, and refer to B05 for details.
For our local $z\sim 0$ sample, we use the NYU Value-added Galaxy
Catalog \citep[VAGC;][]{vagc:05}, based on the second data release of
the SDSS (DR2). The VAGC catalog was used to obtain
\sersic\ parameters, $r$-band half-light radii, and magnitudes. We
then construct a sample of disk-dominated galaxies by requiring
\sersic\ parameter $n<2.5$. We convert the $r$-band half-light radii
to rest-frame $V$-band radii using the conversion derived in B05,
$R_e(V) = 1.011 R_e(r)$. We compute stellar masses from the
(k-corrected) $g$ and $r$-band photometry, using the prescription of
\citet{bell:03}, which relies on a conversion between $g-r$ color and
average stellar mass-to-light ratio. We assume a normalization for
this relation consistent with a \citet{kroupa:01} IMF.

Our high redshift disk sample comes from the GEMS survey
\citep{rix:04,caldwell:06}, which consists of $V_{606}$ and $z_{850}$
imaging over an area of $\sim 900$ arcmin$^2$ with the ACS on HST. The
5$\sigma$ point source detection limit is 28.3 magnitude in the
$V_{606}$ band and 27.1 magnitude in $z_{850}$. The GEMS object
catalog is based on the $z_{850}$ image; for details see
\citet{caldwell:06}. High-accuracy photometric redshift estimates
($\sigma_z/(z+1) \sim 0.02$) are obtained from the ground-based
COMBO-17 survey \citep{wolf:04}. The $R$-band selection limit of
COMBO-17 ($m_R \sim 24$) limits the range of the sample to redshifts
$z\la 1$.  The 17-band photometry of COMBO has also been used to
obtain stellar mass estimates \citep{borch:06}, assuming a Kroupa
IMF. The GEMS main sample consists of almost 8000 galaxies with COMBO
counterparts and redshifts. We fit each galaxy with a \sersic\ profile
and select a disk-dominated sample with good quality fits,
\sersic\ $n<2.5$, and extended light profiles. This sample contains
5664 objects. Typical uncertainties are $\sim 35$\% in $r_{e}$ and
$\sim 0.2$ magnitudes in $m_{z}$ \citep{haussler:06}. The apparent
half-light sizes measured in the observed $z_{850}$ band are converted
to rest-frame V-band using an average color gradient correction based
on a local sample of disk galaxies (see B05). These corrections are
small ($\pm 3$ \%) over the entire redshift range of our sample (the
observed $z_{850}$ band samples the rest-frame $V$-band at
$z\sim0.5$). Where disk scale radii are quoted in this work, we have
obtained them by simply assuming that the measured half-light radius
and the disk scale radius are related by the standard expression for a
pure exponential disk, $r_d = r_e/1.68$.

In order to estimate the completeness of the combined GEMS+COMBO disk
sample, we have performed extensive simulations
\citep{haussler:06,rix:04}. Artificial disks were inserted into blank
sky, and the source detection and fitting software was run on this
image. Poor fits are excluded in the same manner as for the real galaxy
images. We can then calculate the success rate for detecting and
obtaining a good fit for the artificial galaxies, as a function of
apparent effective radius and apparent magnitude. We multiply this
GEMS completeness factor by the (redshift, magnitude, and
color-dependent) probability that the galaxy would be detected and
successfully assigned a redshift in the COMBO survey. Based on these
estimates, B05 argue that GEMS is not surface brightness limited even
in the highest redshift bin, and that the combined GEMS+COMBO sample
is complete down to stellar masses of $10^{10} \msun$. As in B05, we
limit our analysis to galaxies with stellar mass greater than this
value, and we weigh galaxies by the inverse completeness factor in
computing distributions and means. To avoid using galaxies with very
large weights, we exclude objects with a completeness factor smaller
than 0.5. B05 have shown that the average sizes and surface densities
are insensitive to the choice of this limiting completeness factor at
our adopted stellar mass limit.

\begin{figure*} 
\begin{center}
\epsscale{0.7}
\plotone{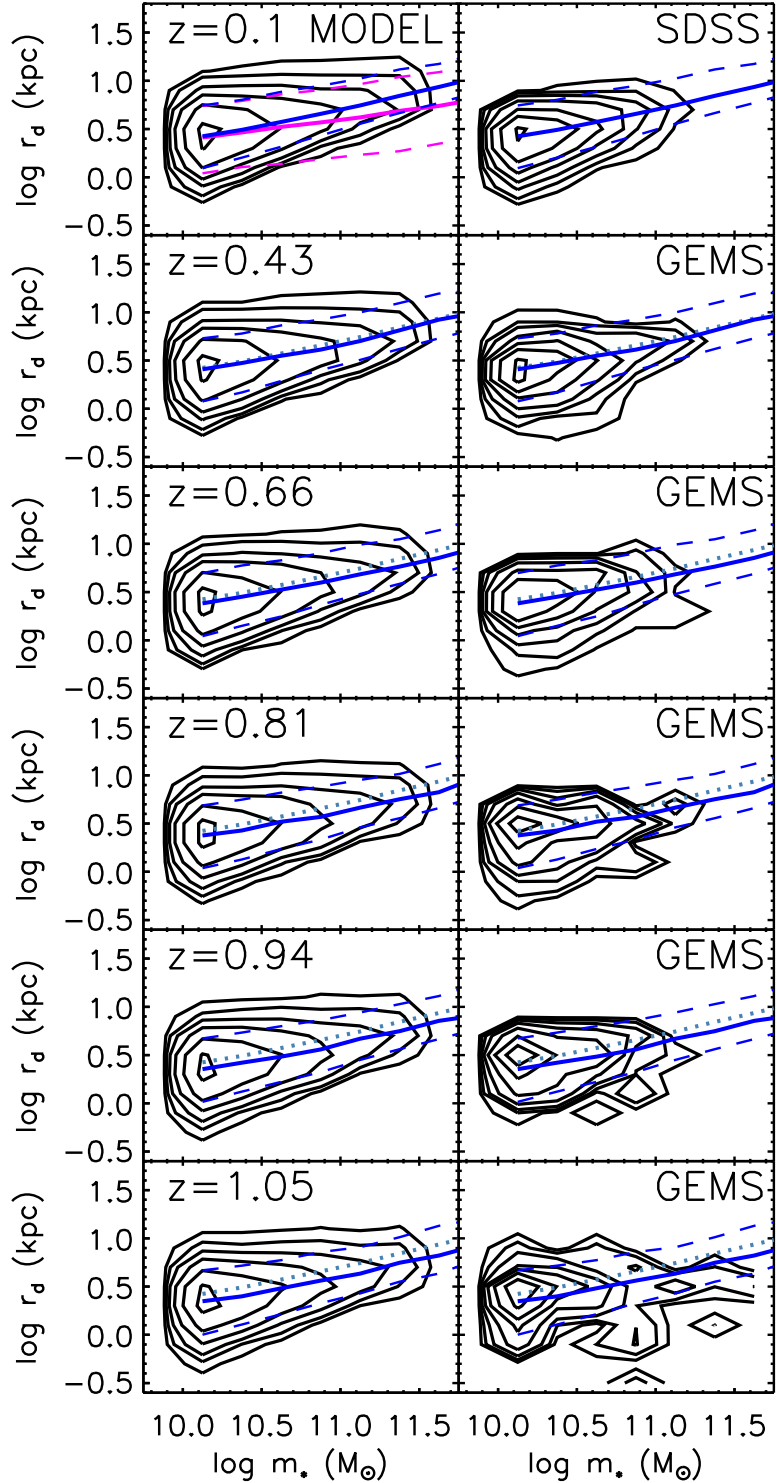}
\end{center}
\caption{\small The relationship between stellar mass and disk scale
  length for a ``local'' sample ($0.001<z<0.2$) and in five redshift
  bins of approximately equal comoving volume from $z=0.1$ to
  $z=1.1$. In the left panels, contours show the NFW model predictions
  for {\it stable} disks. In the right panels, contours show the
  completeness-corrected distributions for the SDSS and GEMS samples
  for the same redshift bins. The left and right panels are normalized
  to the same total number density in each redshift bin. The dark blue
  solid and dashed lines show the median and 10 and 90th percentiles,
  respectively, for the stable {\it model} disks {\it in both columns
    of panels} (models and data). The magenta solid and dashed lines
  in the left $z=0.1$ panel show the median and 10 and 90th
  percentiles for all disks, without any stability criterion
  applied. The gray-blue dotted lines, repeated in each panel, show
  the $z=0.1$ medians and 10 and 90 percentile lines for the stable
  model disks.
\label{fig:sizemstar}}
\epsscale{1.0}
\end{figure*}

\begin{figure*} 
\begin{center}
\plotone{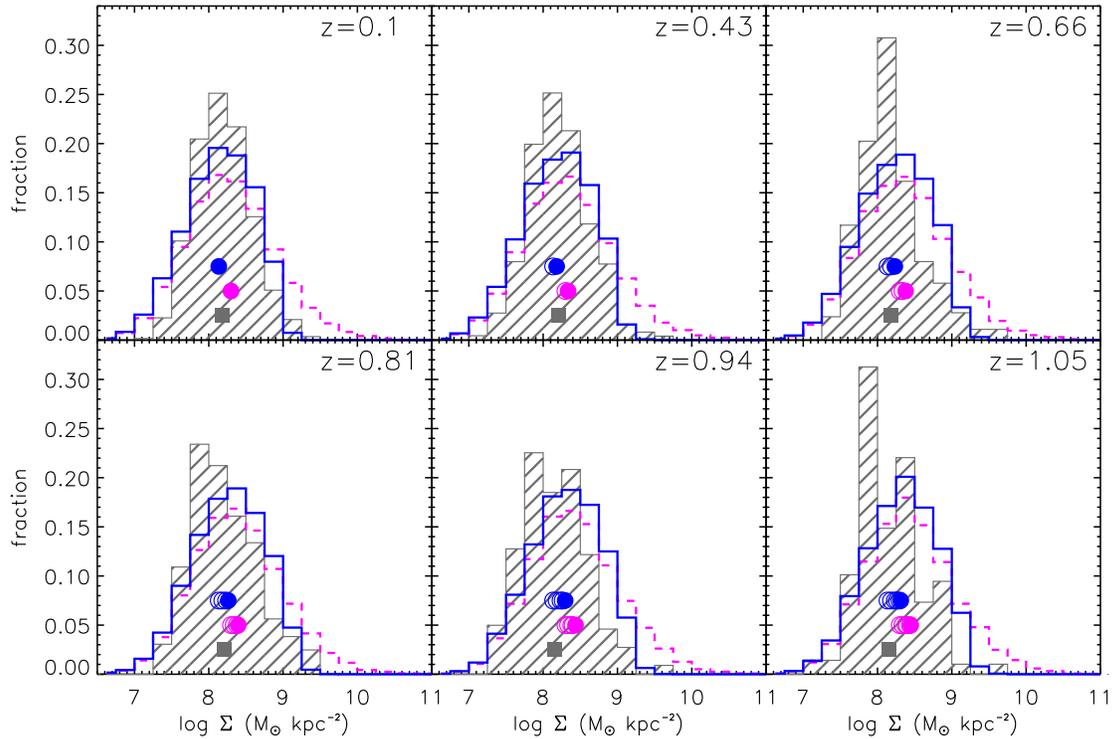}
\end{center}
\caption{\small The distribution of stellar surface densities for
  disks with stellar mass $10^{10} \msun < m_* < 10^{11} \msun$, for a
  ``local'' sample ($0.001<z<0.2$), and in five redshift bins of
  approximately equal comoving volume from $z=0.1$ to $z=1.1$. Shaded
  histograms show the completeness-corrected distributions of SDSS and
  GEMS disks, selected via S\'ersic fits to their radial light
  profiles ($n<2.5$). Solid squares show the mean value of $\log
  \Sigma_*$ derived from these observations.  Magenta dashed lines
  show the distributions for all model disks, and dark blue solid
  lines show the results for ``stable'' model disks only (see text).
  The solid dots indicate the mean of the model distribution for the
  current redshift bin, and the open dots show the means from all
  lower redshift bins. The upper set of dark blue dots are for stable
  disks, and the lower set of magenta dots are for all disks.
\label{fig:sigmadist}}
\end{figure*}

\begin{figure} 
\begin{center}
\plotone{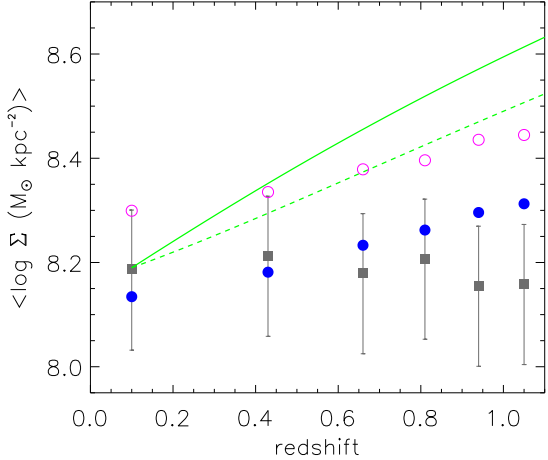}
\end{center}
\caption{\small The redshift evolution of the mean stellar surface
  density. Solid squares show the mean of $\log \Sigma_*$ for SDSS and
  GEMS disks with $10^{10} \msun < m_* < 10^{11} \msun$. Open magenta
  circles show the results for all NFW model disks in this mass range,
  and solid blue dots show the results for stable model disks
  only. The dashed curve shows the evolution in $\Sigma_*$ that we
  would expect if disk size scaled like $r_{200}$, and the solid curve
  shows the evolution for disk sizes that scale like $r_{\rm vir}$ (as
  in the SIS model; see text). The SIS model predicts that disk
  surface densities were considerably higher at $z\sim1$, in conflict
  with the data.
\label{fig:sigmaev}}
\end{figure}

\begin{figure} 
\begin{center}
\plotone{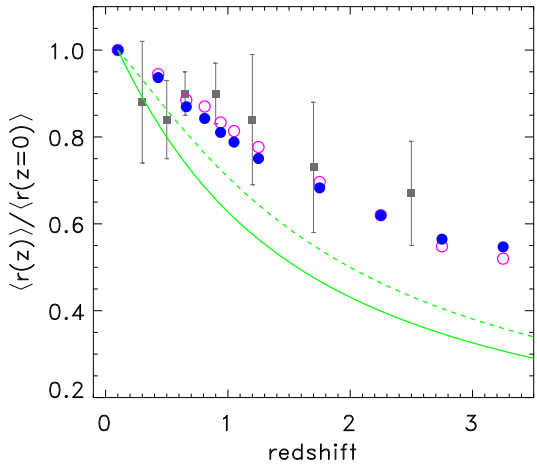}
\end{center}
\caption{\small The redshift evolution of the average size of disks
  with stellar masses greater than $3\times10^{10} \msun$, relative to
  the average size of disks at $z=0.1$. Square symbols with error bars
  show the observational estimates from T06, obtained by combining the
  SDSS, GEMS, and FIRES datasets. Open (magenta) circles show the NFW
  model predictions for all disks. Solid (dark blue) dots show the NFW
  model predictions for stable disks only. The green dashed and solid
  curves show the scaling of $r_{\rm 200}$ and $r_{\rm vir}$
  (respectively) for dark matter halos of fixed mass, as in the SIS
  model. The NFW model predicts more gradual size evolution, in better
  agreement with the observations than the SIS model.
\label{fig:sizeev}}
\end{figure}

\begin{figure} 
\begin{center}
\plotone{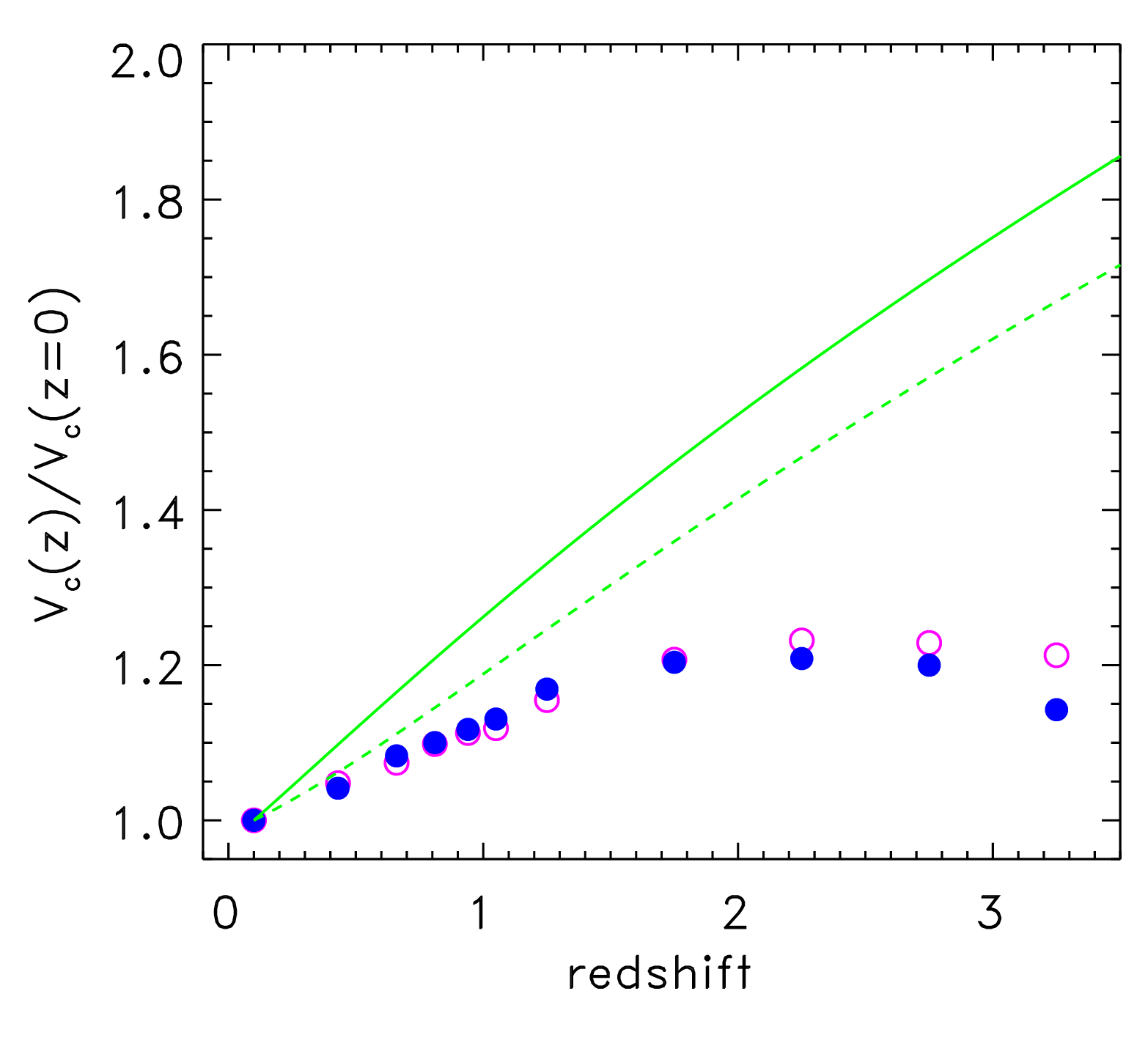}
\end{center}
\caption{\small The redshift evolution of the circular velocity of
  disks with stellar mass 5--$7\times10^{10} \msun$, relative to the
  average circular velocity of such disks at $z=0.1$.  Open (magenta)
  circles show the model predictions for all disks. Solid (dark blue)
  dots show the model predictions for stable disks only. The green
  dashed and solid curves show the scaling of $V_{\rm 200}$ and
  $V_{\rm vir}$ (respectively) for dark matter halos of fixed mass, as
  in the SIS model. The NFW model predicts more gradual evolution in
  rotation velocity, in qualitative agreement with Tully-Fisher
  observations to $z\sim1$ (see text).
\label{fig:vc_ev}}
\end{figure}

\section{Results}
\label{sec:results}

Using the models outlined above, we construct a mock catalog for a low
redshift slice ($0.001 < z < 0.2$) representative of the SDSS sample,
as well as a lightcone from $0.1 < z < 1.1$ with approximately ten
times the area of GEMS ($10 \times 900$ arcmin$^2$), which we divide
into five bins with roughly equal comoving volume ($0.1<z<0.56$,
$0.56<z<0.74$, $0.74<z<0.87$, $0.87<z<1.0$, $1.0<z<1.1$). Following
MMW98, we define ``stable disks'' as those with stability parameter
$\epsilon_m$ above a critical value, and adopt $\epsilon_{m, \rm crit}
= 0.75$ \citep{syer:99}. The stellar mass vs. disk scale length
relations for the six redshift bins from $z\sim0$--1 are shown in
Figure~\ref{fig:sizemstar}, for the mock catalogs on the left and the
observed SDSS and GEMS disk samples on the right\footnote{Note that,
  strictly speaking, the models predict the scale length of the {\it
    baryonic mass} in the disk, while the observed scale lengths are
  for the rest-frame $V$-band light. We do not attempt to correct for
  the known $\sim 20 \%$ difference between these two quantities, since
  we are mainly interested in the redshift evolution. We do note here,
  however, that time-evolving color gradients could therefore change
  our results.}. The observational samples are corrected for
completeness as described in Section~\ref{sec:data}.  

\subsection{The size-mass relation and its evolution to $z\sim1$}

In the left panels of
Fig.~\ref{fig:sizemstar}, the contours show the results for {\it
  stable disks} only. To illustrate the impact of excluding the
unstable disks, the median and 10 and 90th percentiles in size as a
function of stellar mass for all disks, without any stability
criterion applied, are shown in the $z=0.1$ panel. The relative impact
of applying the stability criterion on the size-mass relation and its
scatter is similar in all the redshift bins considered here, so for
clarity we show only the median relation for stable disks in the other
panels. It is clear from the definition of the stability parameter
$\epsilon_m$ that at a given halo mass, disks with larger values of
$f_d$ are more likely to be unstable, and also at a given stellar
mass, more compact disks (low $r_d$) are more likely to be
unstable. Therefore, more massive galaxies are more likely to be
unstable (recall that we assumed that $f_{d}$ increases with
increasing halo mass), and so excluding the unstable (compact) disks
results in a steepening of the slope in the $m_*$--$r_d$ relation at
$m_* \simeq 2-3 \times 10^{10} \msun$. Excluding the unstable disks
also significantly reduces the scatter in disk size at fixed stellar
mass for the massive disks. The fraction of disks deemed stable by our
criterion is nearly constant over the redshift range considered here,
and ranges from 95\% at $z\sim0.1$ to 97\% at $z\sim1$.

Figure~\ref{fig:sizemstar} shows that the NFW model can reproduce both
the observed slope of the size-mass relation and the scatter in size
at a given stellar mass fairly well at $z\sim0$. Our results agree
with those of S03 and other investigations. The median and 10 and 90th
percentiles in disk scale length as a function of stellar mass for the
stable model disks from the low redshift bin are repeated in every
panel, and against these contours we can immediately see that the
model predicts that the average size of disks at fixed stellar mass
has increased by about 15--20\% since $z\sim1$. The GEMS sample is
consistent with no evolution in size at fixed stellar mass. Note that
we have normalized the model and data histograms to the same total
number density in each redshift bin.

An important side issue is that we could see in
Figure~\ref{fig:sizemstar} that the model produces too many massive
disks ($m_* > 10^{11}\msun$), especially at high redshift, compared
with the observational samples. This is not surprising, since we have
assumed that {\it every} dark matter halo contains a single disk
galaxy, which is clearly not realistic, and we know that more massive
halos (which produce the massive disks) are more likely to instead
host an early type galaxy. Therefore, in order to avoid any bias from
the unrealistically massive model galaxies, in
Figures~\ref{fig:sigmadist} and \ref{fig:sigmaev}, we show these
distributions for galaxies with stellar masses in the range $10^{10}
\msun < m_* < 10^{11} \msun$.

Figure~\ref{fig:sigmadist} shows the distribution of stellar surface
densities $\Sigma_*$ for the same six redshift bins from $z\sim0.1$ to
$z\sim1$. We define the stellar surface density as $\log \Sigma_* =
\log m_* - 2 \log r_e - \log (2\pi)$, where $r_e = 1.68 r_d$. From
this figure, we can readily see that if we had not excluded unstable
disks, the model would have predicted a broader distribution of
surface densities than is seen in the data, with a highly skewed tail
to very high densities that are not observed in the disk samples. With
unstable disks excluded, the width of this distribution is reasonably
consistent with the data at all redshifts, though there are hints of
some interesting discrepancies; namely, the observational
distributions appear perhaps a bit narrower and a bit more skewed than
the model predictions.

Figure~\ref{fig:sigmaev} shows the evolution of the average value of
$\log \Sigma_*$ for the same stellar mass range. As seen in the
previous figure, the model predicts significant, but fairly mild,
evolution in $\log \Sigma_*$ over the redshift range considered (about
0.2 dex, or about a factor of 1.5; consistent with $\Sigma_* \propto
r^{-2}_{\rm e}$).  We also show the SIS scalings both for $r_{200}$
and for $r_{\rm vir}$ (see \S\ref{sec:model})\footnote{Note that the
  evolution is somewhat more rapid when the definition $r_{\rm vir}$
  is used, because the virial overdensity used to define the halo,
  $\Delta_{\rm vir}$, evolves with redshift while with the $r_{200}$
  definition it remains constant. Because $\Delta_{\rm vir}$ is larger
  at higher redshift, the halos are smaller in radius.}. Both of these
predict much more dramatic redshift evolution than is observed: as
found in B05, and as seen here, the results from the GEMS analysis are
consistent with no evolution in the average value of $\Sigma_*$ over
the redshift range $0.1\la z \la 1.1$.

B05 found that the 2-$\sigma$ error bars on the average values of
$\log \Sigma$ obtained from bootstrap resampling are $\sim \pm
0.04$--0.1 dex, but this certainly underestimates the possible
systematic errors. For example, the stellar mass estimates may be
systematically incorrect if galaxies have more bursty star formation
histories at high redshift, or the size estimates could be
systematically biased by the increasingly irregular morphologies of
high redshift disks (our fitting simulations assume perfectly smooth
galaxies). We estimate the overall \emph{systematic} uncertainty in
size at fixed stellar mass to be $\sim 30$\%, and show representative
error bars reflecting this in Figure~\ref{fig:sigmaev}. The prediction
of the NFW model is a significant improvement over the SIS scaling,
and it is in acceptable agreement with the GEMS data to $z\sim1$
within these estimated uncertainties.

The surface density of disks is important because nearly all
semi-analytic galaxy formation models assume that the efficiency of
star formation is determined by the disk surface density through the
Kennicutt Law, $\dot{\Sigma}_* \propto \Sigma_{\rm gas}^N$ above a
critical surface density $\Sigma_{\rm crit}$. Predicting disk surface
densities correctly is therefore very important for self-consistent
models of galaxy formation.  Note, however, that many semi-analytic
models \citep[e.g.][]{somerville_primack:99,kauffmann:99,croton:06}
effectively assume that the disk surface density scales according to
the SIS prediction. This will clearly lead to fairly serious
inaccuracies in the predictions of the evolution of the star formation
rates in galaxies in these models.

\subsection{Evolution to higher redshift}

We now briefly explore how the models fare in comparison with the more
limited data available at higher redshift. Figure~\ref{fig:sizeev}
shows the average size of disk galaxies with stellar mass greater than
$3 \times 10^{10} \msun$ predicted by the model, out to $z\sim3$,
compared with the combined results from SDSS, GEMS, and FIRES,
presented by T06. Each set of points is normalized relative to the
average scale length of that sample at $z=0.1$. Results are shown for
all disks, and for ``stable'' disks only. Note that the {\it absolute}
average disk scale lengths for the ``stable'' disks are always larger
than those for the total set of disks, with no stability criterion
applied, however, the redshift evolution of the stable disk sample is
slightly steeper than that of the overall sample out to $z\sim
2$. This is why the stable disk sizes, when normalized relative to the
average size of stable disks at $z=0$, are a little bit lower than the
total disk sample out to $z\sim2$. We again also show the scaling for
halo virial radius at fixed halo mass, which would predict much more
rapid size evolution than is observed. As we have already seen, the
improved model predicts fairly mild evolution in the average disk
sizes out to $z\sim1$, in quite good agreement with the GEMS data. At
higher redshifts $z\sim 2$--3, our model predicts that disks should be
about 60\% as large as they are today at a given stellar mass. This
represents somewhat more evolution than the observational results of
T06 indicate, but is within the quoted error bars.

\subsection{Evolution of the Tully-Fisher relation}

Our models also simultaneously predict the rotation curves of the
disks. It has been shown in previous work that models similar to the
one presented here can reproduce the detailed rotation curves for the
Milky Way and M31 \citep{kzs:02}, and the Tully-Fisher relation
(circular velocity at some fiducial radius vs. luminosity, or
vs. stellar or baryonic mass) at redshift zero
\citep{mo:98,dutton:07,gnedin:06}. A detailed comparison with
observations of rotation velocities and the Tully-Fisher relation is
beyond the scope of this paper, but we briefly present a prediction
for the evolution of the Tully-Fisher zero-point (here expressed as
the average rotation velocity at $\sim$3 disk scale lengths, for disk
galaxies of a fixed stellar mass, $m_* \sim 6 \times 10^{10} \msun$)
in Fig.~\ref{fig:vc_ev}. We find that the disk rotation velocity
$V_{\rm disk} \equiv V_c(r=3 r_s)$ at fixed disk mass evolves much
more gradually than the halo virial velocity $V_{\rm vir}$ at fixed
halo mass. In fact, our model predicts that $V_{\rm disk}$ should
actually remain nearly constant from $z\sim 1.5$--3. Observational
constraints on the time evolution of the Tully-Fisher relation,
particularly in terms of stellar or baryonic mass, remain uncertain,
but \citet{conselice:05} found no evolution in the K-band or stellar
mass Tully-Fisher to $z\sim1$, and \citet{kassin:07} also find no
evolution to $z\sim 1$ in a redefined version of the stellar mass
Tully-Fisher relation\footnote{Instead of using the usual rotation
  velocity, they define a new velocity variable which includes
  contributions from random motions as well as rotation.}.

\section{Discussion and Conclusions}
\label{sec:conclude}

We have shown that a CDM-based model of disk formation produces good
agreement with the observed weak redshift evolution of the disk
size-stellar mass relation from GEMS out to $z\sim1$. This result is
in contrast to the considerably stronger evolution implied by the
assumption often used in the literature, that disk sizes simply scale
in proportion to the dark matter halo virial radii. Similarly, we find
much weaker evolution in the disk rotation velocity at a given disk
mass than we would expect if $V_{\rm disk}$ scaled in proportion to
the halo virial velocity $V_{\rm vir}$.

The more gradual evolution in the ``improved'' model results primarily
from the use of realistic NFW halo profiles, and from the
incorporation of the redshift evolution of the halo concentration-mass
relation predicted by N-body simulations. Recall that in our model, at
fixed halo mass, there are two halo properties and two disk parameters
that determine the structural properties of the disk: halo
concentration \cvir, halo spin $\lambda$, the baryon fraction of the
disk $f_d$ and the fraction of specific angular momentum captured by
the disk materal, $f_j$. We have assumed that the distribution of spin
parameters $\lambda$ for the overall population of dark matter halos
does not change with time, as demonstrated by N-body simulations. We
have also assumed that $f_d$ is a fixed function of halo mass that
does not depend on redshift (though in reality this may not be true),
and we have assumed a fixed value of $f_j=1$. Therefore, there is only
one ingredient left that can possibly be responsible for changing the
predictions for the redshift dependence of disk sizes at fixed mass:
the halo concentration vs. mass relation.

As we have discussed, simulations have shown that the halo
concentration vs. mass relation depends on redshift: the halo
concentration at fixed mass scales as $\cvir \propto (1+z)^{-1}$
\citep{bullock:01a}.  Therefore, a halo of a given mass is less
concentrated at high redshift. This apparent evolution, however, is
really a consequence of the way that halos are assembled in a CDM
universe. Studies of the mass accretion history of halos in
simulations has shown that they have two basic phases of growth: an
early, rapid phase, in which the central density is set, and a second
phase of more gradual accretion \citep{wechsler:02}. The mass within
the characteristic scale radius $r_s$ is assembled during the early,
rapid accretion phase. Afterwards, $r_s$ stays nearly constant, while
$r_{\rm vir}$ increases due to smooth accretion of mass, leading to a
formal decrease in $\cvir \equiv r_{\rm vir}/r_s$.

In Section~\ref{sec:model} we demonstrated that less concentrated
halos produce larger disks because of the lower mass and weaker
gravitational forces in the central parts of the halo, where the disk
forms. Thus the collapse of baryons produces less contraction of the
dark matter profile, and a more extended disk. The trend towards lower
concentrations at earlier times therefore counteracts the decreasing
virial radii. Out to about $z\sim1$, these competing effects nearly
cancel out, leading to the weak evolution in the size-mass relation
that we have shown.

Note that we have repeated our calculations using the ``WMAP3''
cosmological parameters derived by \citet{spergel:06}. Because of the
reduced small scale power in this cosmology, halos form somewhat later
and therefore have lower concentrations for their mass at the present
day. However, once we renormalize our model (by making a minor
adjustment to the normalization of the disk fraction relation $f_d$)
to again reproduce the size-mass relation of present day disks, we
find that the redshift evolution of disk sizes and circular velocities
is nearly unchanged. This is not surprising, as the \emph{redshift
  evolution} of the halo concentration vs. mass relation is nearly
identical to the one we originally adopted.

We see a hint that the evolution predicted by these models is still a
bit stronger than that indicated by the data.  This could be a sign
that one of the other assumptions in our simple model is
incorrect. For example, if the disk baryon fraction $f_{\rm d}$ at a
given halo mass decreases with increasing redshift, this would lead to
shallower evolution and relatively larger disks at high
redshift. Because we have measured the disk sizes in the rest-$V$
band, evolving color gradients could also mask evolution in the true
size of the stellar disk. Alternatively, systematic biases in our
stellar mass and size estimates could be impacting the observational
estimates.

We also see an increasing level of discrepancy at higher redshifts, $z\ga
1.5$. This could be a hint that an entirely different mechanism could
be responsible for setting the sizes of disks at very high
redshift. Mergers between gas-rich disks could result in a new, more
spatially extended disk
\citep{kazantzidis:05,springel_hernquist:05,robertson:06}. This same
scenario could also help to explain the kinematics of Damped
Lyman-$\alpha$ systems, which are difficult to reconcile with the
standard Fall-Efstathiou picture of disk formation \citep{maller:01}.

We also presented a prediction for the redshift evolution of the
average rotation velocity at $\sim 3 r_d$ (close to the maximum of the
rotation curve) for disks of fixed stellar mass $m_* \sim 6 \times
10^{10} \msun$ (i.e., the zeropoint of the Tully-Fisher relation). We
found a similar result as we did for the disk sizes: $\vdisk$
decreases with time, but much more slowly than we would expect if we
assumed that $\vdisk \propto \vvir$. The physical reason for this is
the same as the one we have discussed in relation to the weak size
evolution: halos have lower concentrations at high redshift, and thus
less mass near the center and so lower rotation speeds at the radii
where rotation velocities are measured (near the maximum value of the
rotation curve, $V_{\rm max}$).Thus $V_{\rm max}$ even for pure DM NFW
halos, without accounting for baryons, evolves more slowly than
$\vvir$ \citep{bullock:01a}. In addition, in our model, lower
concentrations lead to less contraction and more extended disks. This
leads to flatter, less ``peaky'' rotation curves, and lower values of
$\vdisk$ relative to a more concentrated halo \citep{mo:98,kzs:02}.

About the same fraction of disks are classified as unstable according
to the condition we adopted ($\epsilon_m<0.75$) over the whole
redshift interval $0 \leq z \leq 3$, and the exclusion of unstable
disks from the sample changes the average size by nearly the same
amount over this interval as well. Therefore, as implemented here,
disk stability does not play a significant role in determining the
relative time evolution of the stellar mass-size relation, although it
is important for reproducing the correct distribution of disk stellar
surface densities. At very high redshifts, $z\ga 2.5$, our model
predicts that the number of unstable disks starts to
increase. However, we do not even know that thin, rotationally
supported disks exist at such high redshifts, so we do not know how
seriously to take this prediction.

While the model presented here represents a significant improvement
with respect to the overly simplified $\lambda r_{\rm vir}$ scaling
commonly used in the literature, it still neglects many important
aspects of disk formation in a hierarchical universe, in particular
the impact of mergers. We have also ignored the possible presence of
spheroids and cold gas in our disk galaxies. As well, the fraction of
baryons in the disk component as a function of halo mass ($f_d$), here
assumed to be a simple deterministic function, almost certainly has a
large scatter and may change systematically with time. We intend to
investigate the predictions of more detailed models, set within
hierarchical merger trees, and including a full treatment of cooling,
star formation, feedback, etc., in a future work, in which we will
also explore the redshift evolution of the disk size function
(Somerville et al., in prep).

\section*{Acknowledgments}
\begin{small}

Based on observations taken with the NASA/ESA {\it Hubble Space
  Telescope}, which is operated by the Association of Universities for
Research in Astronomy, Inc.\ (AURA) under NASA contract NAS5-26555.
Support for the \gems\ project was provided by NASA through grant
number GO-9500 from the Space Telescope Science Institute, which is
operated by the Association of Universities for Research in Astronomy,
Inc., for NASA under contract NAS5-26555. Support for this work also
came from HST Archival Grant AR-10290. SFS acknowledges financial
support provided through the European Community's Human Potential
Program under contract HPRN-CT-2002-00305, Euro3D RTN. EFB is
supported by the DFG's Emmy Noether Program. CW was supported by a
PPARC Advanced Fellowship. SJ acknowledges support from the National
Aeronautics and Space Administration (NASA) under LTSA Grant
NAG5-13063 issued through the Office of Space Science. DHM
acknowledges support from the National Aeronautics and Space
Administration (NASA) under LTSA Grant NAG5-13102 issued through the
Office of Space Science. CH is supported by a CITA National
Fellowship. KJ acknowledges support by the German DFG under grant SCHI
536/3-1.

Funding for the Sloan Digital Sky Survey (SDSS) has been provided by
the Alfred P. Sloan Foundation, the Participating Institutions, the
National Aeronautics and Space Administration, the National Science
Foundation, the U.S. Department of Energy, the Japanese
Monbukagakusho, and the Max Planck Society. The SDSS Web site is
http://www.sdss.org/. The SDSS is managed by the Astrophysical
Research Consortium (ARC) for the Participating Institutions. The
Participating Institutions are The University of Chicago, Fermilab,
the Institute for Advanced Study, the Japan Participation Group, The
Johns Hopkins University, Los Alamos National Laboratory, the
Max-Planck-Institute for Astronomy (MPIA), the Max-Planck-Institute
for Astrophysics (MPA), New Mexico State University, University of
Pittsburgh, Princeton University, the United States Naval Observatory,
and the University of Washington.

\end{small}

\bibliographystyle{apj} 
\bibliography{apj-jour,disksize}

\begin{thebibliography}{67}
\expandafter\ifx\csname natexlab\endcsname\relax\def\natexlab#1{#1}\fi

\bibitem[{{Avila-Reese} {et~al.}(1998){Avila-Reese}, {Firmani}, \& {Hern{\'
  a}ndez}}]{afh:98}
{Avila-Reese}, V., {Firmani}, C., \& {Hern{\' a}ndez}, X. 1998, \apj, 505, 37

\bibitem[{{Barden} {et~al.}(2005)}]{barden:05}
{Barden}, M. {et~al.} 2005, \apj, 635, 959

\bibitem[{{Bell} {et~al.}(2003){Bell}, {McIntosh}, {Katz}, \&
  {Weinberg}}]{bell:03}
{Bell}, E.~F., {McIntosh}, D.~H., {Katz}, N., \& {Weinberg}, M.~D. 2003, \apjs,
  149, 289

\bibitem[{Binney \& Tremaine(1987)}]{binney_tremaine:87}
Binney, J. \& Tremaine, S. 1987, Galactic Dynamics (Princeton, NJ: Princeton
  Univ. Press)

\bibitem[{{Blanton} {et~al.}(2005)}]{vagc:05}
{Blanton}, M.~R. {et~al.} 2005, \aj, 129, 2562

\bibitem[{Blumenthal {et~al.}(1986)Blumenthal, Faber, Flores, \&
  Primack}]{blumenthal:86}
Blumenthal, G., Faber, S., Flores, R., \& Primack, J. 1986, \apj, 301, 27

\bibitem[{{Borch} {et~al.}(2006)}]{borch:06}
{Borch}, A. {et~al.} 2006, \aap, 453, 869

\bibitem[{{Bullock} {et~al.}(2001{\natexlab{a}}){Bullock}, {Dekel}, {Kolatt},
  {Kravtsov}, {Klypin}, {Porciani}, \& {Primack}}]{bullock:01b}
{Bullock}, J.~S., {Dekel}, A., {Kolatt}, T.~S., {Kravtsov}, A.~V., {Klypin},
  A.~A., {Porciani}, C., \& {Primack}, J.~R. 2001{\natexlab{a}}, \apj, 555, 240

\bibitem[{{Bullock} {et~al.}(2001{\natexlab{b}})}]{bullock:01a}
{Bullock}, J.~S. {et~al.} 2001{\natexlab{b}}, \mnras, 321, 559

\bibitem[{{Burstein} {et~al.}(1997){Burstein}, {Bender}, {Faber}, \&
  {Nolthenius}}]{burstein:97}
{Burstein}, D., {Bender}, R., {Faber}, S., \& {Nolthenius}, R. 1997, \aj, 114,
  1365

\bibitem[{Caldwell {et~al.}(2006)}]{caldwell:06}
Caldwell, J.~A.~R. {et~al.} 2006, ApJ, submitted, astro-ph/0510782

\bibitem[{Christodoulou {et~al.}(1995)Christodoulou, Shlosman, \&
  Tohline}]{christodoulou:95}
Christodoulou, D., Shlosman, I., \& Tohline, J. 1995, \apj, 443, 551

\bibitem[{{Conselice} {et~al.}(2005){Conselice}, {Bundy}, {Ellis}, {Brichmann},
  {Vogt}, \& {Phillips}}]{conselice:05}
{Conselice}, C.~J., {Bundy}, K., {Ellis}, R.~S., {Brichmann}, J., {Vogt},
  N.~P., \& {Phillips}, A.~C. 2005, \apj, 628, 160

\bibitem[{{Croton} {et~al.}(2006){Croton}, {Springel}, {White}, {De Lucia},
  {Frenk}, {Gao}, {Jenkins}, {Kauffmann}, {et~al.}}]{croton:06}
{Croton}, D.~J., {Springel}, V., {White}, S.~D.~M., {De Lucia}, G., {Frenk},
  C.~S., {Gao}, L., {Jenkins}, A., {Kauffmann}, G., {et~al.} 2006, \mnras, 365,
  11

\bibitem[{Dalcanton {et~al.}(1997)Dalcanton, Spergel, \&
  Summers}]{dalcanton:97}
Dalcanton, J., Spergel, D., \& Summers, F. 1997, \apj, 482, 659

\bibitem[{{Dekel} \& {Silk}(1986)}]{dekel-silk:86}
{Dekel}, A. \& {Silk}, J. 1986, \apj, 303, 39

\bibitem[{{Dutton} {et~al.}(2007){Dutton}, {van den Bosch}, {Dekel}, \&
  {Courteau}}]{dutton:07}
{Dutton}, A.~A., {van den Bosch}, F.~C., {Dekel}, A., \& {Courteau}, S. 2007,
  \apj, 654, 27

\bibitem[{Efstathiou {et~al.}(1982)Efstathiou, Lake, \&
  Negroponte}]{efstathiou:82}
Efstathiou, G., Lake, G., \& Negroponte, J. 1982, \mnras, 199, 1069

\bibitem[{{Eke} {et~al.}(2001){Eke}, {Navarro}, \& {Steinmetz}}]{ens:01}
{Eke}, V.~R., {Navarro}, J.~F., \& {Steinmetz}, M. 2001, \apj, 554, 114

\bibitem[{Fall \& Efstathiou(1980)}]{fall_efstathiou:80}
Fall, S. \& Efstathiou, G. 1980, \mnras, 193, 189

\bibitem[{{Ferguson} {et~al.}(2004)}]{ferguson:04}
{Ferguson}, H.~C. {et~al.} 2004, \apjl, 600, L107

\bibitem[{Flores {et~al.}(1993)Flores, Primack, Blumenthal, \&
  Faber}]{flores:93}
Flores, R., Primack, J., Blumenthal, G., \& Faber, S. 1993, \apj, 412, 443

\bibitem[{Giavalisco {et~al.}(1996)Giavalisco, Steidel, \&
  Macchetto}]{giavalisco:96}
Giavalisco, M., Steidel, C., \& Macchetto, D. 1996, \apj, 470, 189

\bibitem[{{Gnedin} {et~al.}(2004){Gnedin}, {Kravtsov}, {Klypin}, \&
  {Nagai}}]{gnedin:04}
{Gnedin}, O.~Y., {Kravtsov}, A.~V., {Klypin}, A.~A., \& {Nagai}, D. 2004, \apj,
  616, 16

\bibitem[{{Gnedin} {et~al.}(2006){Gnedin}, {Weinberg}, {Pizagno}, {Prada}, \&
  {Rix}}]{gnedin:06}
{Gnedin}, O.~Y., {Weinberg}, D.~H., {Pizagno}, J., {Prada}, F., \& {Rix}, H.-W.
  2006, ArXiv Astrophysics e-prints

\bibitem[{{Governato} {et~al.}(2004)}]{governato:04}
{Governato}, F. {et~al.} 2004, \apj, 607, 688

\bibitem[{{H{\"a}ussler} {et~al.}(2006)}]{haussler:06}
{H{\"a}ussler}, B. {et~al.} 2006, ApJ, submitted

\bibitem[{{Jesseit} {et~al.}(2002){Jesseit}, {Naab}, \& {Burkert}}]{jesseit:02}
{Jesseit}, R., {Naab}, T., \& {Burkert}, A. 2002, \apjl, 571, L89

\bibitem[{{Jing}(2000)}]{jing:00}
{Jing}, Y.~P. 2000, \apj, 535, 30

\bibitem[{{Kassin} {et~al.}(2007){Kassin}, {Weiner}, {Faber}, {Koo}, {Lotz},
  {Diemand}, {Harker}, {Bundy}, {et~al.}}]{kassin:07}
{Kassin}, S.~A., {Weiner}, B.~J., {Faber}, S.~M., {Koo}, D.~C., {Lotz}, J.~M.,
  {Diemand}, J., {Harker}, J.~J., {Bundy}, K., {et~al.} 2007, \apjl, 660, L35

\bibitem[{Kauffmann(1996)}]{kauffmann:96}
Kauffmann, G. 1996, \mnras, 281, 475

\bibitem[{{Kauffmann} {et~al.}(1999){Kauffmann}, {Colberg}, {Diaferio}, \&
  {White}}]{kauffmann:99}
{Kauffmann}, G., {Colberg}, J.~M., {Diaferio}, A., \& {White}, S.~D.~M. 1999,
  \mnras, 303, 188

\bibitem[{{Kazantzidis} {et~al.}(2005)}]{kazantzidis:05}
{Kazantzidis}, S. {et~al.} 2005, \apjl, 623, L67

\bibitem[{{Klypin} {et~al.}(2002){Klypin}, {Zhao}, \& {Somerville}}]{kzs:02}
{Klypin}, A., {Zhao}, H., \& {Somerville}, R.~S. 2002, \apj, 573, 597

\bibitem[{{Kroupa}(2001)}]{kroupa:01}
{Kroupa}, P. 2001, \mnras, 322, 231

\bibitem[{{Lilly} {et~al.}(1998)}]{lilly:98}
{Lilly}, S. {et~al.} 1998, \apj, 500, 75

\bibitem[{Lowenthal {et~al.}(1997)}]{lowenthal:97}
Lowenthal, J. {et~al.} 1997, \apj, 481, 673

\bibitem[{Macci\'o {et~al.}(2006)Macci\'o, Dutton, van~den Bosch, Moore,
  Potter, \& Stadel}]{maccio:06}
Macci\'o, A., Dutton, A., van~den Bosch, F., Moore, B., Potter, D., \& Stadel,
  J. 2006, /mnras, in press, astro-ph/0608157

\bibitem[{{Maller} \& {Dekel}(2002)}]{maller-dekel:02}
{Maller}, A.~H. \& {Dekel}, A. 2002, \mnras, 335, 487

\bibitem[{{Maller} {et~al.}(2001){Maller}, {Prochaska}, {Somerville}, \&
  {Primack}}]{maller:01}
{Maller}, A.~H., {Prochaska}, J.~X., {Somerville}, R.~S., \& {Primack}, J.~R.
  2001, \mnras, 326, 1475

\bibitem[{{Mao} {et~al.}(1998){Mao}, {Mo}, \& {White}}]{mao:98}
{Mao}, S., {Mo}, H.~J., \& {White}, S.~D.~M. 1998, \mnras, 297, L71

\bibitem[{Mo {et~al.}(1998)Mo, Mao, \& White}]{mo:98}
Mo, H., Mao, S., \& White, S. 1998, \mnras, 295, 319

\bibitem[{{Natarajan}(1999)}]{natarajan:99}
{Natarajan}, P. 1999, \apjl, 512, L105

\bibitem[{{Navarro} {et~al.}(1997){Navarro}, {Frenk}, \& {White}}]{navarro:97}
{Navarro}, J.~F., {Frenk}, C.~S., \& {White}, S.~D.~M. 1997, \apj, 490, 493

\bibitem[{{Navarro} \& {Steinmetz}(2000)}]{navarro_steinmetz:00}
{Navarro}, J.~F. \& {Steinmetz}, M. 2000, \apj, 538, 477

\bibitem[{{Navarro} \& {White}(1994)}]{navarro_white:94}
{Navarro}, J.~F. \& {White}, S.~D.~M. 1994, \mnras, 267, 401

\bibitem[{{Peebles}(1969)}]{peebles:69}
{Peebles}, P.~J.~E. 1969, \apj, 155, 393

\bibitem[{Pizagno {et~al.}(2006)}]{pizagno:06}
Pizagno, J. {et~al.} 2006, preprint, astro-ph/0608472

\bibitem[{{Ravindranath} {et~al.}(2004)}]{ravindranath:04}
{Ravindranath}, S. {et~al.} 2004, \apjl, 604, L9

\bibitem[{{Rix} {et~al.}(2004)}]{rix:04}
{Rix}, H.-W. {et~al.} 2004, \apjs, 152, 163

\bibitem[{{Robertson} {et~al.}(2006){Robertson}, {Bullock}, {Cox}, {Di Matteo},
  {Hernquist}, {Springel}, \& {Yoshida}}]{robertson:06}
{Robertson}, B., {Bullock}, J.~S., {Cox}, T.~J., {Di Matteo}, T., {Hernquist},
  L., {Springel}, V., \& {Yoshida}, N. 2006, \apj, 645, 986

\bibitem[{{Robertson} {et~al.}(2004){Robertson}, {Yoshida}, {Springel}, \&
  {Hernquist}}]{robertson:04}
{Robertson}, B., {Yoshida}, N., {Springel}, V., \& {Hernquist}, L. 2004, \apj,
  606, 32

\bibitem[{{Sargent} {et~al.}(2006){Sargent}, {Carollo}, {Lilly}, {Scarlata},
  {Feldmann}, {Kampczyk}, {Koekemoer}, {Scoville}, {et~al.}}]{sargent:06}
{Sargent}, M.~T., {Carollo}, C.~M., {Lilly}, S.~J., {Scarlata}, C., {Feldmann},
  R., {Kampczyk}, P., {Koekemoer}, A.~M., {Scoville}, N., {et~al.} 2006, ArXiv
  Astrophysics e-prints

\bibitem[{{Shen} {et~al.}(2003)}]{shen:03}
{Shen}, S. {et~al.} 2003, \mnras, 343, 978

\bibitem[{{Sheth} \& {Tormen}(1999)}]{sheth_tormen:99}
{Sheth}, R.~K. \& {Tormen}, G. 1999, \mnras, 308, 119

\bibitem[{{Simard} {et~al.}(1999)}]{simard:99}
{Simard}, L. {et~al.} 1999, \apj, 519, 563

\bibitem[{Somerville \& Primack(1999)}]{somerville_primack:99}
Somerville, R. \& Primack, J. 1999, \mnras, 310, 1087

\bibitem[{{Sommer-Larsen} {et~al.}(1999){Sommer-Larsen}, {Gelato}, \&
  {Vedel}}]{sommer_larson:99}
{Sommer-Larsen}, J., {Gelato}, S., \& {Vedel}, H. 1999, \apj, 519, 501

\bibitem[{{Spergel} {et~al.}(2006){Spergel}, {Bean}, {Dor{\'e}}, {Nolta},
  {Bennett}, {Dunkley}, {Hinshaw}, {Jarosik}, {et~al.}}]{spergel:06}
{Spergel}, D.~N., {Bean}, R., {Dor{\'e}}, O., {Nolta}, M.~R., {Bennett}, C.~L.,
  {Dunkley}, J., {Hinshaw}, G., {Jarosik}, N., {et~al.} 2006, ArXiv
  Astrophysics e-prints

\bibitem[{{Springel} \& {Hernquist}(2005)}]{springel_hernquist:05}
{Springel}, V. \& {Hernquist}, L. 2005, \apjl, 622, L9

\bibitem[{{Syer} {et~al.}(1999){Syer}, {Mao}, \& {Mo}}]{syer:99}
{Syer}, D., {Mao}, S., \& {Mo}, H.~J. 1999, \mnras, 305, 357

\bibitem[{{Trujillo} {et~al.}(2004)}]{trujillo:04}
{Trujillo}, I. {et~al.} 2004, \apj, 604, 521

\bibitem[{{Trujillo} {et~al.}(2006)}]{trujillo:06}
---. 2006, \apj, 650, 18

\bibitem[{{van den Bosch}(2000)}]{vandenbosch:00}
{van den Bosch}, F.~C. 2000, \apj, 530, 177

\bibitem[{{Wechsler} {et~al.}(2002){Wechsler}, {Bullock}, {Primack},
  {Kravtsov}, \& {Dekel}}]{wechsler:02}
{Wechsler}, R.~H., {Bullock}, J.~S., {Primack}, J.~R., {Kravtsov}, A.~V., \&
  {Dekel}, A. 2002, \apj, 568, 52

\bibitem[{{Weil} {et~al.}(1998){Weil}, {Eke}, \& {Efstathiou}}]{weil:98}
{Weil}, M.~L., {Eke}, V.~R., \& {Efstathiou}, G. 1998, \mnras, 300, 773

\bibitem[{{Wolf} {et~al.}(2004)}]{wolf:04}
{Wolf}, C. {et~al.} 2004, \aap, 421, 913

\end{thebibliography}

\end{document}